\documentclass[12pt]{iopart}

\usepackage{graphicx,color}

\RequirePackage[normalem]{ulem} %DIF PREAMBLE
\begin{document}
\title[Runaway Electron Control in FTU]{Runaway Electron Control in FTU}

\author{ D Carnevale$^1$,  B Esposito$^2$, M Gospodarczyk$^1$, L Boncagni$^2$,
M Sassano$^1$, S Galeani$^1$, D Marocco$^2$, L Panaccione$^2$, O Tudisco$^2$,  
W Bin$^5$,  C Cianfarani$^2$, G Ferr\`o$^1$,  G Granucci$^5$, A Gabrielli $^1$,
C Maddaluno$^2$,    J R Mart\`in-Sol\`is $^4$ , Z. Popovic$^4$, F Martinelli$^1$,  
G Pucella$^2$, G Ramogida$^2$, M Riva$^2$,  and FTU Team$^3$}

\address{$^1$ Dipartimento di Ingegneria Civile ed Informatica DICII, Universit\`a di Roma, 
Tor Vergata, Via del Politecnico 1, 00133 Roma, Italy. }
\address{$^2$ ENEA Unit\`a Tecnica Fusione, C.R. Frascati, Via E. Fermi 45, 
00044-Frascati, Roma, Italy.}
\address{$^3$ See the appendix of P. Buratti et al., Proceedings of the 24th 
IAEA Fusion Energy Conf., San Diego, USA, 2012}
\address{$^4$ Universidad Carlos III de Madrid, Avenida de la Universidad 30, 
28911-Madrid, Spain}
\address{$^5$ Istituto di Fisica del Plasma, Consiglio Nazionale delle Ricerche (CNR), Milan, Italy}

\ead{daniele.carnevale@uniroma2.it}

\begin{abstract}
Experimental results on the position and current control of disruption generated runaway electrons (RE) 
in FTU are presented.
 A scanning interferometer diagnostic has been used to analyze the 
 time evolution of the RE beam radial position
and its instabilities.
Correspondence of the interferometer time traces, radial profile reconstructed via magnetic 
measurements and 
fission chamber signals are discussed. 
New RE control algorithms, which define in real-time updated plasma
current and position references, have been tested in two experimental scenarios 
featuring disruption generated RE plateaus. 
Comparative studies among 52 discharges with disruption generated RE beam plateaus 
are presented in order to assess the effectiveness
of the proposed control strategies as the RE beam interaction 
with the plasma facing components is reduced while the current is ramped-down.
\end{abstract}


\vspace{2pc}

\submitto{\NF}
%

\maketitle

\section{Introduction}

A crucial challenge towards a safe and efficient operation of ITER
consists in the need of reducing the dangerous 
effects of runaway electrons (RE) during disruptions \cite{lehnen14}.  RE are 
considered to be potentially intolerable for ITER  when exhibiting currents 
larger than 2MA.
One of the most popular strategies to address this task is based mainly on RE 
suppression by means of Massive Gas Injection (MGI) of High-Z noble gas 
before the thermal quench (TQ), which possesses the additional advantage 
of reducing the localized heat load. However, MGI leads to long recovery 
time, requires effective disruption predictors, and may lead to hot tail RE generation \cite{tail}
or high mechanical loads if the Current Quench (CQ) does not occur in
a suitable time interval \cite{lehnen14,Putvinski, HolmanDisruption}.
Nevertheless, in the circumstances in which such suppression strategy is not
effective, for instance due to a delayed detection of the disruption and/or
to a failure of the gas valves or of disruption avoidance techniques ($e.g.$ ECRH)
\cite{avoidance}, an alternative strategy consisting of 
the RE beam energy and population dissipation by means of a 
RE active beam control may be pursued, as noted in \cite{REshift,Lukash,ITERpcs_14}. 
Alternative mitigation techniques exploit magnetic (resonant) fluctuations/perturbations to displace RE; 
their effect on RE beam dissipation have been studied in \cite{lehnenPerturb,Papp_11,Matsuyama_14}.
Resonant magnetic perturbation techniques can be also used at the CQ to prevent large avalanche effects. 
However they require specific active coils that are not available at FTU.\\

The method proposed in this paper achieves stabilization of a disruption generated 
RE beam by minimizing its interaction with the plasma facing components (PFC). The
RE energy dissipation is obtained by reducing the RE beam current via
the central solenoid (inductive effects). Similar techniques have been investigated in 
DIII-D \cite{REshift}.
In particular, the focus here is on those RE that survive the CQ.
When the RE beam position is stabilized, further techniques, 
not studied in the present paper, such as high-Z gas injection to increase 
RE beam radiative losses could be exploited. \\
  
In the last years, experiments on RE active control have been carried out in DIII-D, 
Tore Supra, FTU, JET, and initial studies have been carried out also 
at COMPASS \cite{Vlainic}. 
In Tore Supra attempts of RE termalization via MGI (He) 
have been investigated \cite{SaintLaurent_13}. 
%Thermalization of 1 MeV electron into 1.7 1020 m-3 electron densities requires 30 ms
%This value extrapolates to half a second for 15 MeV
%Thus an active control of RE regime is required. Together
%with a beam position control, the RE current must also be sustained to maintain its value
%inside the operational domain of the plasma control system
In DIII-D disruptions have been induced by injecting either Argon pellets or MGI
while the ohmic coil current feedback has been left active to maintain constant current 
levels or to follow the desired current ramp-down \cite{REshift}. DIII-D also  studied 
the current beam dissipation rate by means of MGI  with a final termination 
at approximatively 100kA  \cite{Hollmann_13}. Similar results on MGI mitigation of RE
have been obtained at JET \cite{lehnenJET}.
The present work goes along similar lines but RE beam 
dissipation is obtained only by inductive effects, $i.e.$ via central solenoid as in \cite{REshift}, 
combined with a new 
dedicated tool of the Plasma Control System (PCS). This scheme yields a RE beam current ramp-down
and position control. Effectiveness of the novel approach is measured in 
term of reduced interaction of highly energetic runaways with the PFC. Furthermore,
as in \cite{Hollmann_13}, we consider the RE beam radial position obtained by the
CO$_2$/CO scanning interferometer, showing that is also in agreement with
neutron diagnostics and the standard real-time algorithm based on magnetic
measurements that estimate the plasma boundary.

A brief list of  FTU diagnostics correlated with this work is given below.
Further details are given in \cite{DiagnosticsFTU}.  

\textbf{Fission Chamber (FC):} a low sensitivity $^{235}$U 
fission chamber manufactured by Centronic, with a coating of 30 $\mu$g/cm$^2$ of $^{235}$U  
operated in pulse mode at 1 ms 
time resolution and calibrated with a $252$Cf source. This diagnostic 
is essential in the analysis of the sequence of events occurring during the RE current plateau phase
since the standard Hard X-ray (HXR) and neutron monitors are typically constantly saturated after the CQ. 
During the RE plateau phase this detector measures photoneutrons and photofissions 
induced by gamma rays with energy higher than 6 MeV 
(produced by bremsstrahlung of the RE interacting with the metallic plasma facing components).

%\textbf{Neutron camera (NC): }
%The FTU NC has six Lines Of Sight (LOS), viewing the plasma from a lower vertical port; 
%each LOS is equipped with a liquid scintillator detector (2" diameter x 2" thick NE213) 
%coupled to a photomultiplier (PMT) and an embedded $^{22}$Na
%source for calibration purposes. The detector is capable of $n/\gamma$ 
%discrimination and the digital acquisition has been optimized for 
%RE studies.
%The neutron camera provides count rate
%of the Hard-X rays (HXR) larger than 100 KeV produced by in-plasma 
%bremsstrahlung of RE \cite{marocco}.

%\textbf{Fast Electron Bremsstrahlung (FEB):} it is a camera, with 5 ms
%integration time, which detects HXR emission in the energy range
%20-200 keV by means of cadmium telluride semiconductors 
%with several chords of view.

\textbf{Soft-x (SXR):} the multichannel bolometer detects $x$ rays emitted 
at the magnetic center of the toroidal camera (major radius equal to 0.96 m) 
in the range of 5eV to 10keV. Within this range also RE collisions with 
plasma impurities can be detected.

\textbf{Hard-x (HXR):} the X-rays are monitored by two systems:
\begin{itemize}
\item[a)] NaI scintillator detector sensitive to hard-x  
rays with energy higher than 200keV mainly emitted by RE 
hitting the vessel (labeled as HXR in the figures). 
\item[b)] The NEU213 detector sensitive both to neutron and 
to gamma rays and cross calibrated with a BF3 neutron detector in 
discharges with no RE \cite{esposito03}. This detector  is used to 
monitor the formation of RE during the discharge, however at the disruption
and during the RE plateau its signal is 
usually saturated and therefore the gamma monitoring is replaced by the 
fission chamber.
\end{itemize}

\textbf{Interferometer:} 
the CO$_2$/CO scanning interferometer can provide the number of electrons 
measured on several plasma vertical chords (lines of sight, LOS) intercepting 
the equatorial plane at different radii ranging from 0.8965 m to 1.2297 m 
with a sampling time of 62.5 $\mu s$. Detailed information related
to mounting position and specific features are given in \cite{interfco2}. 

\textbf{MHD sensors:} the amplitude of the Mirnov coil signal \cite{cesidio}
considered is directly related to helical deformations of the plasma resulting from MHD 
instabilities, having in most cases n=1 (m=2) toroidal (poloidal) periodicity.

\section{Control strategies}\label{sec:controlstrategy}

\begin{figure}[h]
\begin{center}
\resizebox{0.6\columnwidth}{!}{\includegraphics{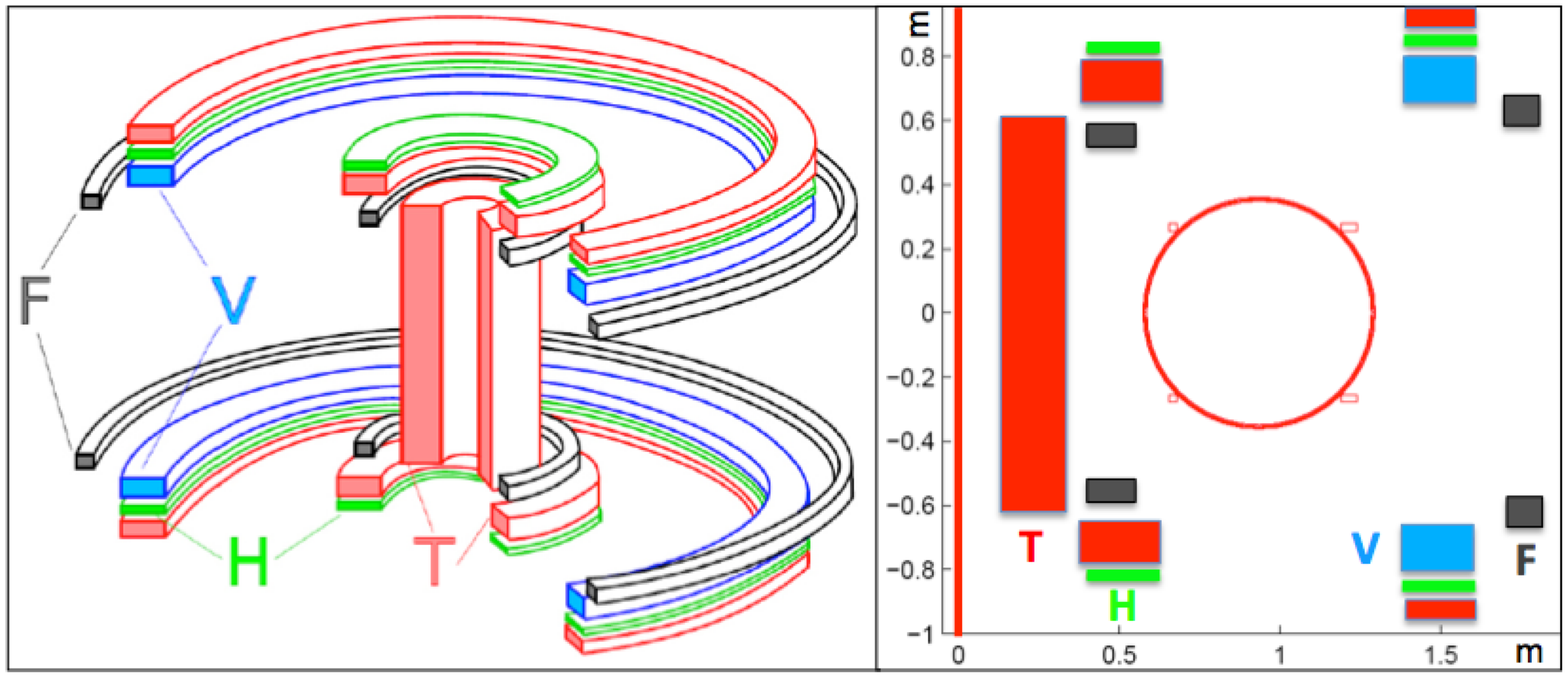}}
\caption{The active coils at FTU: T allows to control the plasma current,
V and F the plasma column radial movements and elongation 
while H the plasma column vertical position.} 
\label{fig:coils} 
\end{center}
\end{figure}

Dedicated FTU plasma discharges have been performed to test two novel  
real-time (RT) RE control algorithms, named PCS-REf1 and PCS-REf2. Such algorithms
have been implemented within the framework of the FTU plasma control system (PCS)
for position and $I_p$ ramp-down control of disruption-generated RE. 
The active coils used to control the position and the current 
of the plasma are shown in Figure \ref{fig:coils}. 
The PCS of FTU, extensively described in \cite{FEDpcs}, exploits the current flowing 
within the T coil, called the central solenoid, to impose the plasma current 
via inductive effect. The  T coil current $I_T$ is regulated via a feedback control 
scheme based on a Proportional-Integral-Derivative (PID) regulator,
which is driven by the plasma current error plus a preprogrammed signal.
The horizontal position of the plasma is controlled by means of an additional
PID regulator that is fed with the horizontal position error. The latter error is 
obtained by on-line processing of a series of pick-up coil signals to determine the
plasma boundary (last closed magnetic surface) which is compared
along the equatorial plane to the desired plasma internal 
$R_{\mathrm{int}}$ and external $R_{\mathrm{ext}}$ radii, 
see \cite{FTUhorizontal,Ariola,FTUmigration,FTUnew} for further details.  
The current flowing in the F coil ($I_F$), by geometrical construction, 
allows us also to modify the plasma elongation $\hat e$. 
%In standard configurations, 
%the elongation satisfies $\hat e \approx k_0 - k_1 I_F/I_p$, 
%with $k_0=1.03$ and $k_0=-4.61$. 
The current on the V coil ($I_V$), 
which allows us to produce a vertical field similar to F but with a slower rate of change, 
is modified by a specific controller named  Current Allocator \cite{Boncagni_13} 
in order to change at run-time  $I_F$ and maintain  
unchanged the vertical field. In such a way, the plasma radial position 
is left unchanged and meanwhile it is possible to steer the value 
$I_{F}$ away from saturation levels. The current redistribution (reallocation) 
between $I_F$ and $I_V$ is performed by the Current 
Allocator at a slower-rate than the changes imposed on $I_F$ by the 
PID regulator (PID-F) for plasma horizontal stabilization (two time scale 
feedback system \cite{khalil}).\\

The PCS safety rules impose that whenever the HXR signal takes value above a given 
safety threshold  (0.2) for more than 10 ms, indication that harmful
RE are present, the discharge has to be shut-down.
In the previous shut-down policy the $I_p$ reference was exponentially 
decreased down to zero and the desired $R_{int}$ and $R_{ext}$
where left unchanged.\\
The new controller \textit{PCS-REf1} has been specifically designed for RE beam dissipation 
and comprises two different phases. In the first phase, specific algorithms 
described in \cite{Boncagni_13} are employed to detect the CQ and the 
RE beam plateau by processing the $I_p$ and the HXR signal.
At the same time, the Current Allocator steers the values of $I_F$ away from saturation
limits, to ensure that a larger excursion
is available for the control of the RE beam position. 

In the second phase, once the RE beam event has been detected (CQ or HXR level), 
the $I_p$ reference is ramped-down
in order to dissipate the RE beam energy by means of the central solenoid. 
In particular,  a scan of the initial values and slope 
of the updated $I_p$ reference for RE suppression (current ramp-down), 
that substitutes the original $I_p$ reference when the RE beam is detected, have been 
performed.  
At the same time the desired (reference) external radius $R_{\mathrm{ext}}$ is reduced
linearly with different slopes down to predefined constant values.
However, the updated $R_{\mathrm{ext}}$ 
reference is such that, below 1.1 m, it is constrained to 
be not smaller than  $R_{\mathrm{ext}}-0.03$ m to avoid 
large position errors that might induce harmful oscillations of the RE beam 
due to the action of the PID-F position controller.
The $R_{ext}$ has been reduced in order to compensate 
for a large outward shift of the RE beam, 
hence to preserve the low field side vessel from RE beam impacts. 
The reduction of the external plasma radius reference 
can be considered the way of finding the RE beam radial position 
that provides minimal RE beam interaction with PFC, similar 
findings have been discussed in \cite{REshift} and the RE beam position
with minimal PFC interaction is called the ``safe zone''. In all the experiments,
the internal radius $R_{int}$ is not changed since we are operating in (internal)
limiter configuration. Nevertheless, the control system has the objective to 
maintain the plasma within the desired horizontal and 
vertical radii, avoiding the plasma impact with the vessel (both side). \\   
A second novel controller \textit{PCS-REf2} has been designed with the same objective 
of RE beam control and energy  suppression. The main difference between 
this second controller and PCS-REf1 consists in an alternative profile for 
$R_{\mathrm{ext}}$, when the latter is ramped-down. 
In this case the updated reference of $R_{\mathrm{ext}}$ is 
ramped-down to a specific constant value, within the range
$[1.11,\,1.13]$ m, associated to low level of the FC signal during the 
experiments where PCS-REf1 was active, plus a small time-varying term. 
This small time-varying term constrained to belong to the set $[-0.04,0.04]$ m, is 
computed in real-time by processing the measured 
$R_{\mathrm{ext}}$ and FC signals according to 
the extremum seeking technique (similar to a gradient algorithm 
discussed in \cite{ex_ftu,ex_ftu_cdc}) in order to minimize the real-time FC signal. 
Furthermore, the $I_p$ ramp-down slope selected for 
PCS-REf2 is about three times smaller than PCS-REf1.
Note that due to the current amplifiers limitations the control system is not 
expected to be effective in position and $I_p$ current ramp-down control 
within $25-30$ ms of the CQ detection.

%\redt{PCS-REf3 has been tested with the second scenario.}
%\item  low-prefill, Current Input Allocator active, Ne gas injection 
%about 30ms before a step of  the plasma desired current from 360kA up to 500kA, 
%density reference as before.

\section{Experimental results}

%\footnote{The rate of success to obtain RE beam plateau 
%lasting more than 10ms injecting Ne is less than 15\% at FTU. }.

\subsection{Experimental scenarios}
The new RE control architecture has been applied in low-density plasma discharges. In a \textit{first 
scenario} a significant RE population is generated during the $I_p$ ramp-up/flat top at
360kA by selecting low gas prefill and (low) density reference of $1.5E-19$  m$^{-3}$, 
followed by an injection of Ne gas to induce a disruption. The sudden variation of the resistivity and 
the increased loop voltage at the disruption accelerate the pre-existing RE population 
and lead, in some cases\footnote{Formation of the RE plateau depends upon a number of 
factors at TQ and CQ that cannot be easily controlled.}, to the formation of a RE current plateau which is the target 
scenario of these experiments. 
Note that this is not a method to create runaways but to turn an existing seed population 
of RE in a runaway plateau at the disruption. The discharge is run with an initial 
low gas prefill in such a way that early in the discharge ramp up a runaway population 
is established.
The controller PCS-REf1 has been tested in the above
scenario whereas the controller PCS-REf2 has been tested in a \textit{second scenario}
that differs with respect to the previous one in terms of an extremely low gas prefill 
that causes spontaneous disruption during the $I_p$ ramp-up. This leads, in some cases, 
to a RE plateau. Note that since Ne gas is not injected to induce disruptions, the 
$Z_{\mathrm{eff}}$ is generally less than the first scenario.

\begin{figure}[h]
\begin{center}
\resizebox{0.85\columnwidth}{!}{\includegraphics{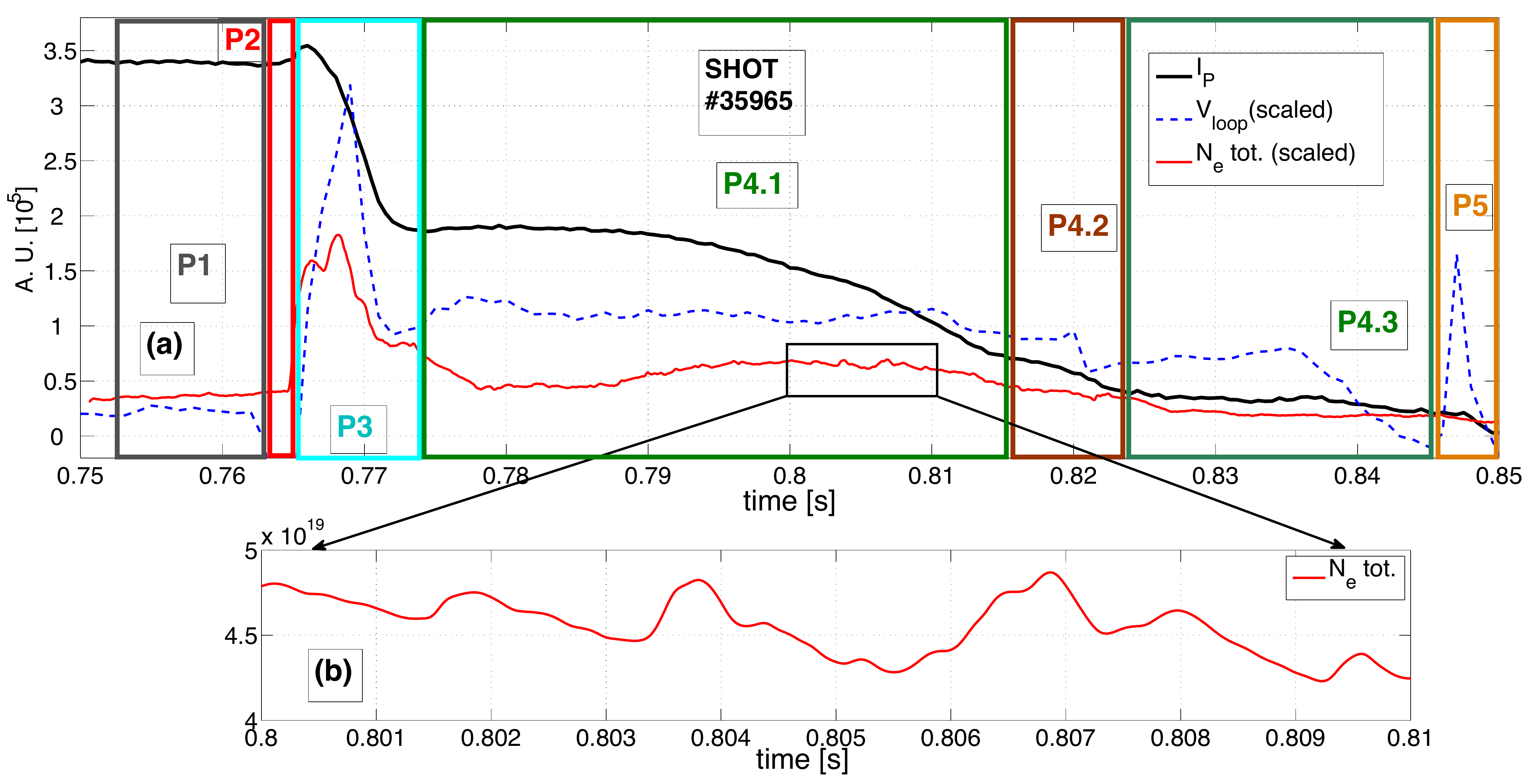}} % 1_UPDATE_LINES.pdf}}
\caption{Pulse $\#35965$: plasma/RE current (black),  total number of 
electrons obtained by the scanning interferometer using the 
Gaussian interpolation profiles (red solid), loop voltage (dashed blue). 
Different pulse phases are highlighted from P1 to P5 and the time trace of
the total number of electrons is magnified within the time window 
$[0.80,\,0.81]$ in the subplot (b).}  \label{fig:IplLunga_1}
\end{center}
\end{figure}

\subsection{RE beam analysis} 

The characterization of the different phases of a disruption 
with the generation of a RE plateau is given in Figure \ref{fig:IplLunga_1}
for the discharge $\#35965$, which may be considered a typical instance  of the first scenario. 
After  Ne gas injection, the plasma density slightly increases 
during the pre-disruptive phase P1 (grey box).  
%that initially contribute to the 
%plasma current inducing its small increment. 
The TQ  (phase P2, orange box), lasting few milliseconds (1-2ms),  
in which the plasma confinement is lost and the thermal 
energy is released to the vessel combined with the high electrical field, 
produces a large increase of the electron density. 
The CQ phase P3 (green box) follows: it is 
characterized by a sudden drop of the plasma current and a high self-induced
parallel electric field ($V_{\mbox{\tiny loop}}$) that further accelerates
the preexisting RE and possibly increases their number.
If the RE beam survives during the CQ to 
collisionality drag, loss of position control
\footnote{The loss of 
the RE beam during the CQ typically occurs toward 
the inner wall since the vertical field, produced by the active coils,  is not quickly reduced 
to match the fast variation of the CQ.}, MHD induced expulsion (to mention only few 
RE loss phenomena), then the RE plateau phase (P4) is started.
In this specific RE scenario, generally the latter phase P4 can be in turn divided 
into three sub-phases:  during phase P4.1 the RE beam current exponentially 
replaces a large fraction of the ohmic $I_p$ current (see \cite{Hollmann_13}, this process starts 
 with the onset of the CQ); subsequently part of such current can be lost due 
to instabilities (P4.2), while the rest of the beam can survive (further plateau in phase P4.3) 
before the final loss (phase P5). 
\begin{figure}[h]
\begin{center}
\resizebox{1.05\columnwidth}{!}{\includegraphics{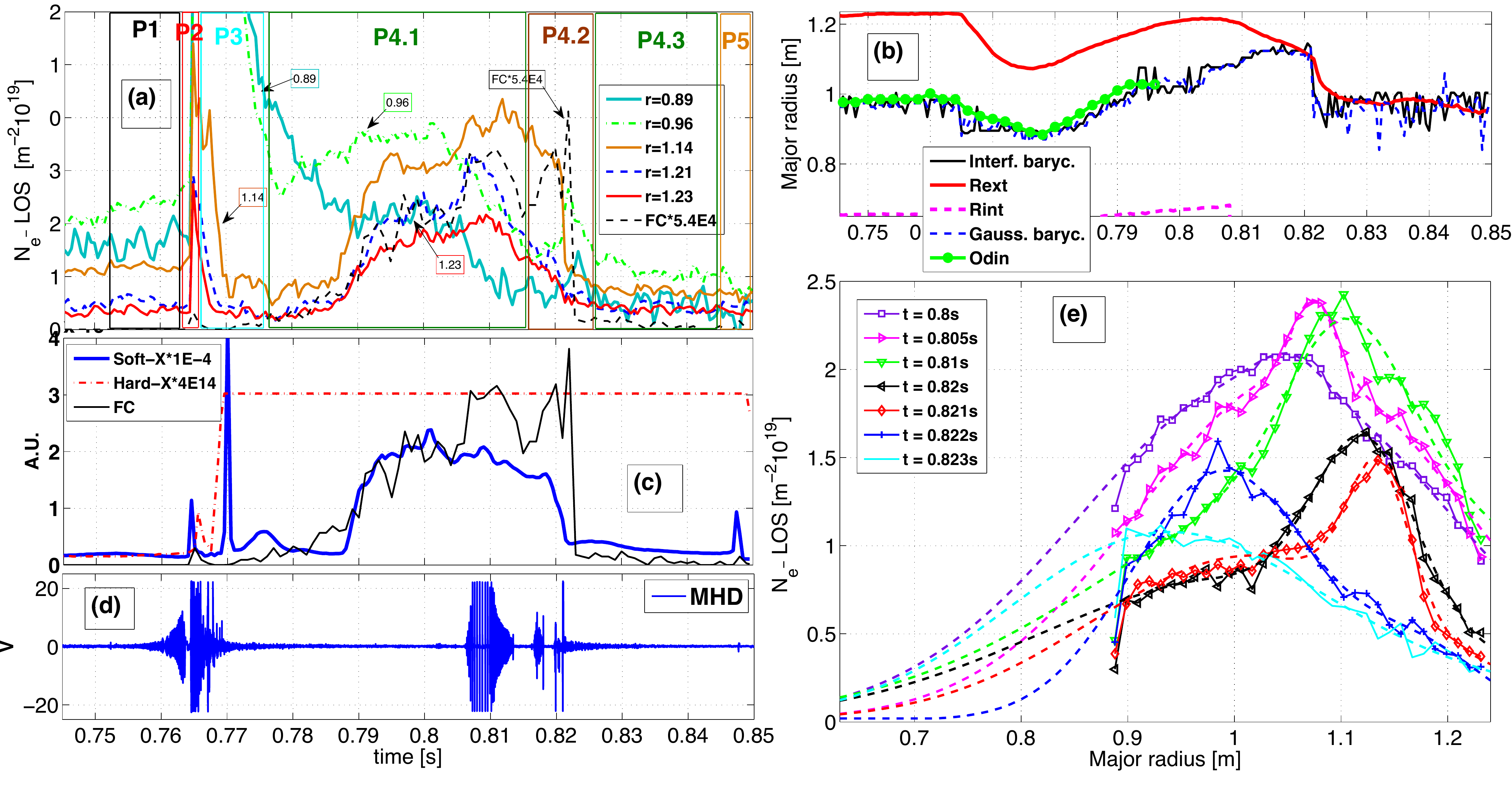}}
\caption{Pulse $\#35965$:   
(a) some of the interferometer LOS at different radii 
compared with the scaled FC signal (black dashed); 
(b) external $R_{\mathrm{ext}}$ (red solid) and internal 
$R_{int}$ (pink dashed) estimated plasma radii at equatorial plane, 
major radius of the density profile highest peak (black, interferometer
barycenter) and the one of the Gaussian functions (blue dashed, Gaussian 
barycenter), magnetic axes reconstructed via ODIN equilibrium code
\cite{FEDpcs} (green solid with circle marks, available up to 0.795 s);  
(c) scaled SXR central line of sight ($R = 0.935$ m, blue) and 
HXR (red) compared with FC counts (black);
(d) Mirnov coil related to MHD activity;
(e) density line integrals at different times (solid) and the 
fitted Gaussian functions (dashed).
} 
\label{fig:35965_tot} 
\end{center}
\end{figure} 

In Figure \ref{fig:IplLunga_1}, beside the time traces
of the plasma/RE beam current (solid black) and  loop voltage (dashed blue), the time trace
of the total number of electrons estimated by fitting 
the scanning interferometer data with a Gaussian function (solid red) are shown. Note that the 
vertical lines of sight of the CO$_{2}$ interferometer are placed only in the central and 
lower field side of the torus (from 0.8965 m to 1.2297 m). We have fitted  
the LOS electrons radial profile with Gaussian functions, whose
parameters have been obtained exploiting least square algorithms, in order to estimate
also the electron density for major radius belonging to the range $[0.63, 0.89]$ m 
(the LOS at 0.8965 m is affected by large measurement noise and has been 
neglected). Example of fitted Gaussian functions on experimental 
data are given in Figure \ref{fig:35965_tot}.(e). The total number of electrons have 
been estimated integrating, with respect to the toroidal angle, the number of electrons 
(evaluated by the Gaussian density radial profile) falling into poloidal plane vertical strips of 
0.5cm radial width (numerical discretization), and summing up all the contributions from 
$R=0.63$ (low-field limiter) up to $R=1.23$ (high-field limiter).\\
       
In order to obtain an estimate of the number of RE
we consider a time interval in which the loop voltage $V_{\mathrm{loop}}$ 
induced by the central solenoid is larger than $5$V. In fact in this case
a decrease in the plasma current $I_p$ 
cannot be associated to the suppression 
 induced by central solenoid but it is essentially due to the RE confinement loss. 
 Furthermore, we restrict the analysis
to the time interval where the interferometer signal indicates
a shift of the electrons toward the (low field-side) poloidal limiter,
$i.e.$ toward the portion of the torus where  the 
interferometer LOS are present.\\
Under such conditions, which are met within the time interval $[0.8,\,0.81]$ s
in discharge  $\sharp 35965$ shown in Figure \ref{fig:IplLunga_1}, 
it is possible to infer that most likely the $I_p$ reduction is directly related 
to the loss of RE (that carry most of the current) in the low field-side 
of the vacuum chamber. Moreover, the FC detector reveals that (a
percentage of) RE have energy higher than about 6 MeV.
In the time interval $[0.8,\,0.81]$ s, letting 
$v_{||,RE} \approx \cos(\overline{\theta}) v_{\mbox{tot,RE}}$ where $\overline{\theta}$
is the RE beam mean pitch-angle, $v_{\mbox{tot,RE}}\approx c$ for $6$ MeV RE, then
by using current drop $\Delta I_p=50kA$, major radius of the RE beam centroid 
(Gaussian barycenter) $R_{re}\approx 1.07$ m, drop of electron density 
$\Delta N_e  =5.6E18$ it is possible to obtain a rough estimate
of the ratio between RE and cold \textit{lost} electrons by means of
\begin{eqnarray}\label{eq:nre}
\frac{n_{re}^{\mathrm{lost}}}{n_{\mathrm{tot}}^{\mathrm{lost}}}\approx \frac{\Delta N_{re}}{\Delta N_{\mathrm{tot}}}
\approx \frac{\Delta I_p2\pi R_{re}}{e^{-} v_{||,\mathrm{re}} \Delta N_e},
\end{eqnarray}
belonging to the range $[1.3,\,2.5]$E$-3$,  assuming $\overline\theta\in[20^\circ\,,60^\circ]$
as reported in \cite{esposito03}.
The percentage of the RE with respect to background plasma
estimated using the total number of electrons $N_e$ within the same 
time interval by using 
\begin{eqnarray}\label{eq:nreok}
\frac{n_{re}}{n_{\mathrm{tot}}}\approx 
\frac{I_p 2\pi R_{re}}{e^{-} v_{||,\mathrm{re}} N_e},
\end{eqnarray}
belongs to the range $[2.6,\,5.2]$E$-3$,  assuming $\overline\theta\in[20^\circ\,,60^\circ]$.
Similar ratios have been found in DIII-D \cite{Hollmann_13} and Tore-Supra \cite{SaintLaurent_13}.
It is interesting to note the similar ratios between lost and beam runaway electrons.
Moreover since the cold plasma closely moves together 
with the RE beam, the RE should be subject to a further 
mechanism of energy loss due to its magnetic interaction with the cold plasma
that, as the synchrotron radiation \cite{ramon} effect, would drain/limit energy 
of the RE beam. 

%\dancom{}{It is interesting to note that the ratio of the non-lost RE
%is similar (double) with respect to the lost ones. This would mean that the magnetic
%coupling of energetic electrons leaving the beam would influence cold electrons 
%trajectories in such a way that also the background plasma is lost.}\\

\noindent In the following we discuss the RE beam
dynamics in the phase P4.1-P4.3 looking at the signals of Fig. \ref{fig:35965_tot}. 
The reader can also refer to the density LOS profile versus time  
shown in the left graph of  Fig.\ref{fig:35965_3D}.\\
The trends of the LOS in the time interval between the RE plateau onset (about  0.775 s) 
and 0.82 s, depicted in Figure \ref{fig:35965_tot}.(e), 
show that the cold plasma column is 
moving toward the low-field side: correspondingly the FC signal increases due to the increased 
RE loss on the outer limiter (low-field side, phase P4.1).  \\
In Figure \ref{fig:35965_tot}.(e)  a large outer shift and an 
extremely peaked profile of the cold plasma column 
can be seen at approximatively 0.81 s. 
Such large shift towards the low-field side is 
caused by the control system that handles $I_F$ and $I_V$ to obtain a 
centered RE beam (0.96 m) as shown by an analysis on the equilibrium conditions obtained by
a simple formula given in \cite{radialFTU}, also confirmed by the more 
sophisticated CREATE NL tool \cite{artaserse}. 
Anyway the presence of a large outward shift of the RE beam orbits
is indeed a well known feature of the RE beam \cite{REshift}. Furthermore,   the vertical magnetic 
field produced by the active coils (F and V) cannot be responsible 
for the high peaked profile (same analysis as above). 
Therefore, the RE beam that magnetically confines 
the cold electrons  is responsible for 
the peaked LOS profiles having a large radial shift. 
This fact is in agreement with \cite{Hollmann_13} where it 
is shown that the RE beam is enveloped by the cold plasma and this 
supports the use of the interferometer data to estimate 
the barycenter of the RE beam.

In Figure \ref{fig:35965_tot}.(b) we have reported the time traces 
of $R_{\mathrm{int}}$ and $R_{\mathrm{ext}}$, that are the internal 
and external plasma radii evaluated via magnetic measurements of the pick-up 
coils (see \cite{FEDpcs}). In the same subplot, we have depicted the 
\textit{interferometer barycenter}, i.e. the major radius corresponding to the 
highest peak of the interferometer density profile,
and the \textit{Gaussian barycenter} that is the major radius corresponding to
the highest peak of the Gaussian functions that are very similar to the 
magnetic axes depicted in green solid line with circle marks obtained by using the 
ODIN algorithm, an equilibrium code
that uses magnetic measurements \cite{FEDpcs}. For the considered discharge
$\#35965$ of Fig. \ref{fig:35965_tot} the ODIN algorithm 
converges up to 0.795 s.\\ 
These signals allow to estimate the centroid of the RE beam and it can be seen that if the plasma 
is not highly peaked (before 0.805 s as shown in the subplot (e)) and it is not
in the high-field side where CO$_2$ LOS are not present (after 0.823 s), 
trends of magnetically reconstructed plasma external radii and the interferometer
barycenter are very similar (this fact generally holds 
 for RE beam plateau at FTU). 
%As discussed in \cite{smith_06,smith_09},  peaked RE beam current profiles
%may cause tearing modes to become unstable. This might explain the increased MHD activity
%when at the same time the LOS are highly peaked at 0.805 s as shown in 
%the subplot \ref{fig:35965_tot}.(e).
The corresponding increment of the FC signal amplitude is the mark that a greater number of RE,  
displaced  from the RE beam by the MHD activity, hit the vessel.
At 0.821 s a particularly intense MHD activity, registered 
as a sudden spike in the subplot \ref{fig:35965_tot}.(d), associated also to 
highly peaked LOS profile in the subplot (e), displaces a large fraction of the 
RE beam towards the toroidal (inner wall) limiter as shown by the 
inward movement of the cold plasma in the subplot \ref{fig:35965_tot}.(e) (phase P4.2). 
Consequently, there is a large spike also in the FC at 0.822 s.
The displacement of RE towards the toroidal limiter   
induces a drop of the loop voltage, shown in Figure \ref{fig:IplLunga_1}. 
This voltage drop, measured by the coil placed within the inner limiter, 
is induced by flux increment produced by the 
inward displacement of the RE beam. A detailed  study of 
the loop voltage sudden drop at time $0.821$ s, not reported here, has been carried
out confirming the RE beam shift towards the inner wall. 
Detailed studies on this phenomenon have been carried out in \cite{loopVoltage}.

%Afterwards, the outward peaked 
%density line integral profiles indicate also a loss of cold electrons toward the 
%poloidal wall, associated to the decreasing of the interferometer LOS signal with 
%r=1.21 m. The difference between the Soft-X and FC signals 
%(subplot (e)) from t=0.805 s onwards is due to the combination of the RE beam
%outward shift and RE loss.
% unlike what seem to happen in DIII-D \cite{Hollmann_13},  where it is assumed a 
%constant profile along the RE plateau to proceed with a tomographic inversion of 
%the interferometer and SXR emissivity profile to infer RE beam spatial position. \\
% the cold electrons are attracted by
%REs due to standard Amp\`ere's law (attractive magnetic 
%forces between electrons flowing around the torus on the same direction that 
%can be approximated as parallel wires with current flowing on the same direction)
%and further studies are needed. 
%As shown in the subplot (b) of Figure \ref{fig:35965_tot},
%the external plasma radius $R_{\mathrm{ext}}$ can be considered as an estimate 
%of the outer RE beam radius when the density profile are not heavily peaked. 

\begin{figure}[h]
\begin{center}
\resizebox{1.05\columnwidth}{!}{\includegraphics{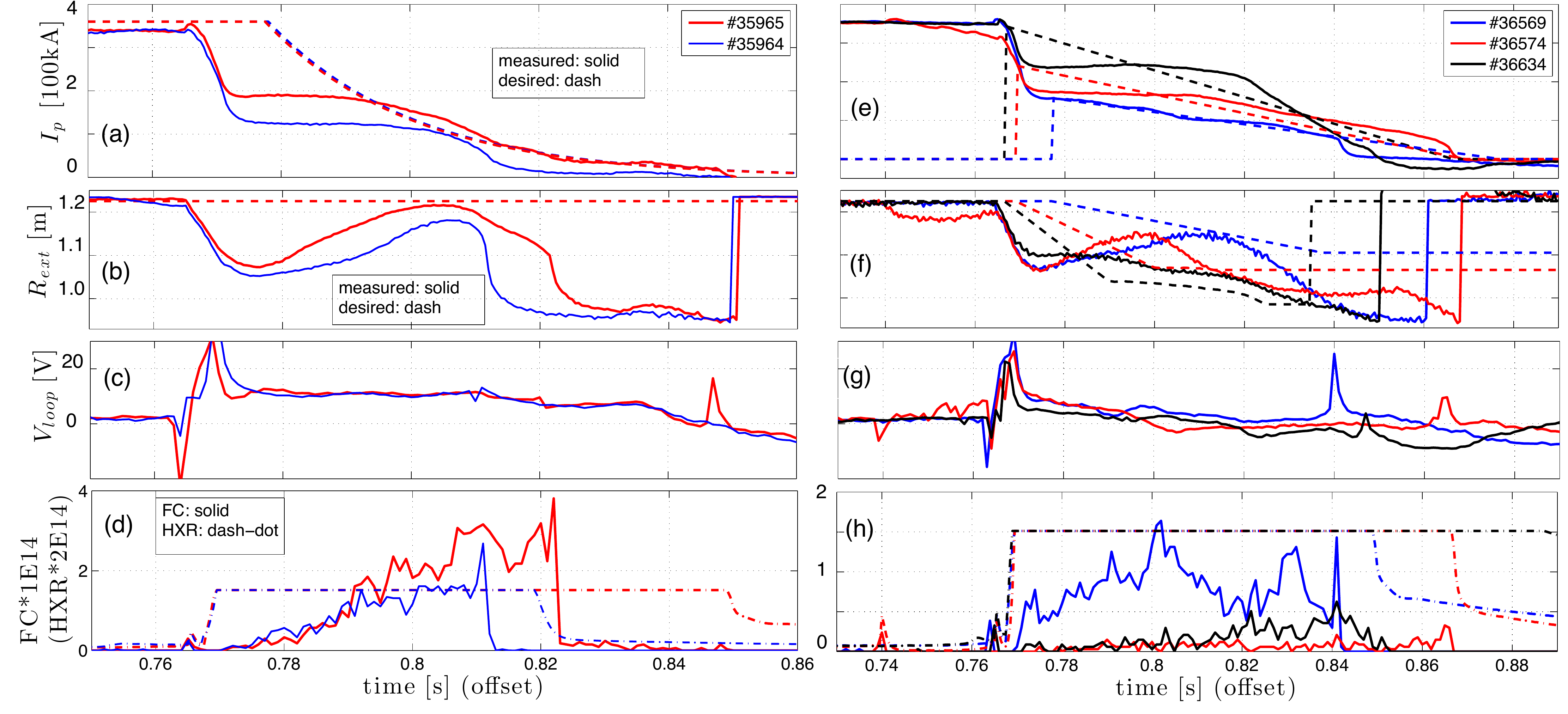}}
\caption{First scenario: Comparison between experimental results  
with the old RE shut-down policy ($\#35965$, $\#35964$) in the left column, 
and the PCS-REf1 controller ($\#36569$, $\#36574$, $\#36634$) in the right column.} 
\label{fig:control_neon} 
\end{center}
\end{figure} 

\subsection{RE position and current control performances}

\begin{figure}[h]
\begin{center}
\resizebox{1.05\columnwidth}{!}{\includegraphics{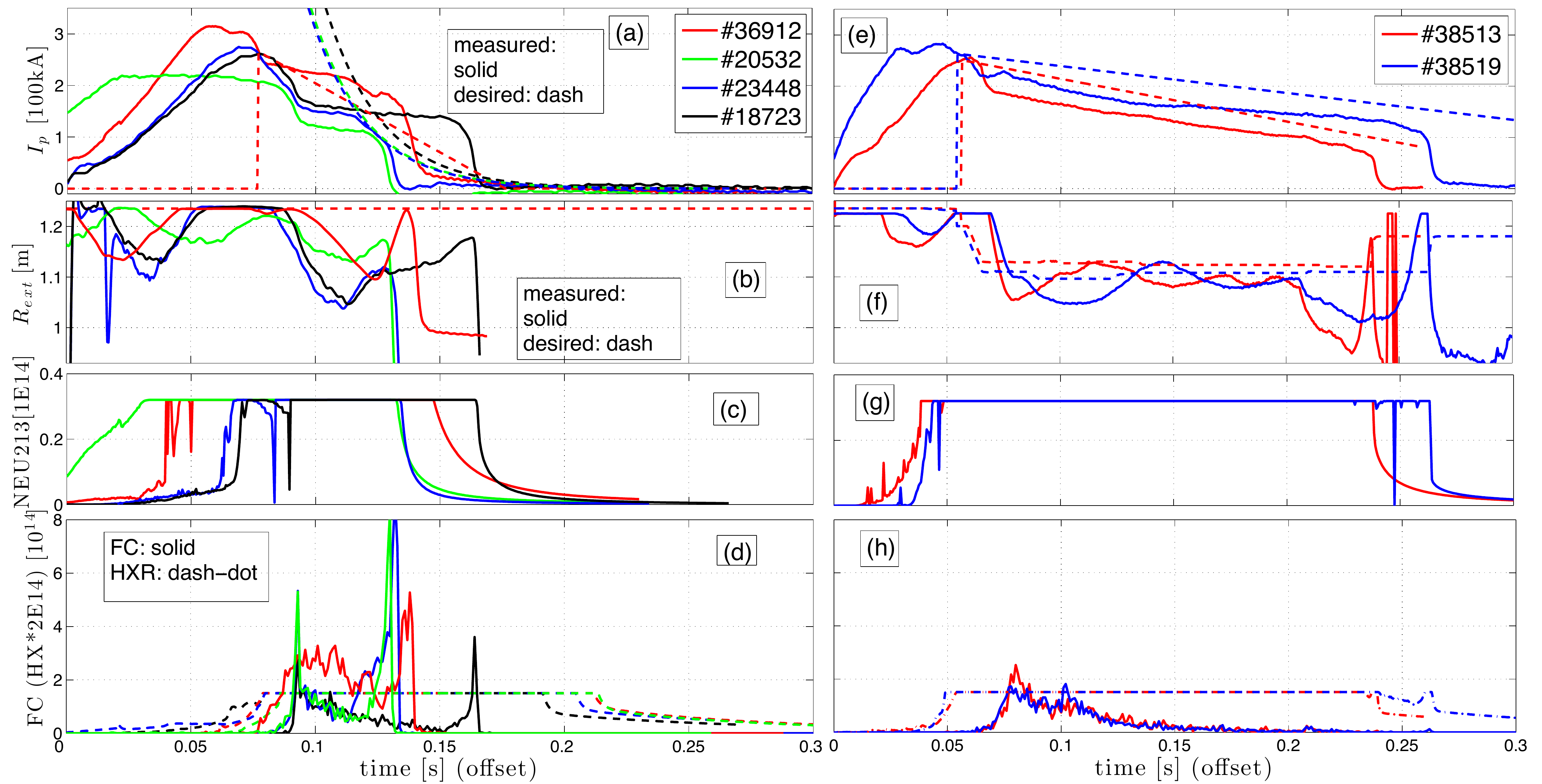}}
\caption{Second scenario: Comparison between experimental results  
with the old shut-down policy ( $\#20532$, $\#23448$,  $\#18723$) 
and only a current ramp-down without the redefinition of the $R_{ext}$ reference 
nor the use of the current allocator ($\#36912$) in the left column. 
Two pulses obtained with PCS-REf2 controller ($\#38513$, $\#38519$) are shown 
in the right column.} 
\label{fig:control_spontaneous} 
\end{center}
\end{figure}

Experimental results of the first and second scenario are  
shown in the Figures \ref{fig:control_neon} and  \ref{fig:control_spontaneous}, respectively. 
In the left column subplots of the two figures the results of the previous control policy are reported.   
The plasma current $I_p$  reference 
at flat top is 360kA, except for the shots $\#20532$, $\#23448$,  $\#18723$
that is equal to $500$ kA.  
In the new policies the old plasma current reference (exponentially
decreased down to zero whenever the HXR signal is above 0.2 for more than 10 ms) 
is substituted by a linear ramp-down as shown by the dashed lines in 
in the subplots (a).
Before the new reference (dashed lines) sets in, the previous one is used. 
In the experiments where the new controller PCS-REf1 has been activated, 
the switching time of the  plasma  current ramp-down has been changed by modifying 
specific parameters of the CQ detector among different pulses, whereas the 
different slopes have been set directly modifying the controller configuration file. 
The alternative plasma current reference allows to define an updated plasma current error that is
fed to the PID-T controller (the PID regulator used for current control in the T coil). 
Then, the PID-T acts on a current amplifier 
in order to change the current flowing within the central solenoid,
$I_T$,  yielding the RE beam current suppression by induction.  
 
Subplots (b, f) of Figures \ref{fig:control_neon} and \ref{fig:control_spontaneous} 
show the estimated (solid) $R_{\mathrm{ext}}$ via magnetic measurements and 
the new  $R_{\mathrm{ext}}$  desired value (dashed). 
In the old policy $R_{\mathrm{ext}}$ was kept constant and equal to 1.23 m, whereas in the 
new PCS-REf1 it is changed dynamically at run-time as discussed 
in Section \ref{sec:controlstrategy}. 
The loop voltages are depicted in the subplots (c, g) of Figure \ref{fig:control_neon} 
whereas the time traces of the NEU213 x-ray monitor are provided in Figure 
\ref{fig:control_spontaneous}. In the subplots (d ,h) the FC (solid) 
and HXR (dashed) signals (the latter, scaled by a factor 2E14, saturates at $1.5\cdot 2E14$) are shown. \\  
The new $I_p$ ramp-down induces a lower 
$V_{\mbox{\tiny loop}}$ in comparison to old shots 
(see  Figure \ref{fig:control_neon}.(c, g)), 
due to the action of the control system that 
modifies the rate of the current in the central solenoid coil,
hence reducing the energy transferred by the central solenoid to the 
RE and consequently reducing also the radial  shift. 
On the contrary, in the old policy, the effort
of the PCS to recover the flat-top $I_p$ value induced large voltages 
that increased the RE energy and consequently also their outward shift. 
At the same time, the smaller $R_{\mathrm{ext}}$ of the new reference contributes 
to the reduction of the RE beam interaction with the  low-field side wall.
Since the discharges of the first scenario (pre-disruption level of RE and the 
FC signal during the CQ) are almost comparable, the improvements in terms 
of reduced FC signal compared to discharge $\#35965$ and $\#35964$
without PCS-REf1 are quite evident. It is interesting
to note the large improvement for the shots $\#36574$ and $\#36634$, also with respect
to the $\#36569$, when the external radius is reduced of more than $10\%$. \\
Data obtained in the first scenario have been processed to determine a suitable 
constant $R_{\mathrm{ext}}$ reference associated to minimal FC signal values. 
The values found are in the range $R_{\mathrm{ext}}\in [1.11,\, 1.13]$ m
against the standard 1.23 m. 
This reduction is in agreement with the RE beam outward shift $\Delta R_{RE}$ 
given by the approximated formula \cite{REshift}
\begin{eqnarray}
\Delta R_{RE} \approx \frac{\overline q W_{RE}}{ecB},
\end{eqnarray}
where $\overline q$ is the averaged safety factor 
$ q\in[2,17]$, 
$W_{RE}\in[6,20]$ MeV is the RE energy, $B=6 T$ is the toroidal field yielding 
$\Delta R_{RE}\in [2,12.5]$ cm.
This new constant values have been used for the controller PCS-REf2, tested in a second scenario,
whose results are shown in Figure \ref{fig:control_spontaneous}.
In the latter case also the real-time FC signal is exploited in order to slightly
modify $R_{\mathrm{ext}}$ and minimize the FC signal: 
see the dashed lines in Figure \ref{fig:control_spontaneous}.(f) that 
suddenly ramp down to 1.11 m and 1.13 m for $\#38519$ and $\#38513$, respectively, 
and then slightly changes in time. Although the number of the available discharges 
in the second scenario was not sufficient to optimally tune the gains of the 
extremum seeking policy,  the results are  encouraging. 

Discharges of the second scenario characterized by a sudden increase of 
the FC signal at CQ have been found on the FTU database and some of them 
have been reported in the left column of Fig. \ref{fig:control_spontaneous} . 
The discharges with active RE beam control
 $\#38513$ and $\#38519$ show a reduction of the FC signal down to zero while 
the $I_p$ is slowly ramped-down and the reference of $R_{ext}$ 
is reduced. On the contrary the discharges  $\#20532$, $\#23448$, $\#18723$, 
without the active control, show a substantial
increase of the FC signal slightly before 
and during the final loss. 
In the discharges  $\#38513$ and $\#38519$
the two hard x-rays monitors NEU213 and HXR shown in 
the figure \ref{fig:control_spontaneous} (c,g) are saturated from the 
CQ throughout the $I_p$ ramp-down indicating that
energetic RE are present. In the discharge $\#38519$ at the end of the current ramp-down,
about 30 ms before the final loss,  the NEU213 have small drops below the saturation value.
In the left column of Fig. \ref{fig:control_spontaneous} it is shown the shot
$\#36912$ that is an example of shot where only a ramp-down current is 
performed and where the current allocator is not active. It can been seen
that after the CQ there is the usual compression against the inner wall
but then, due to the control system that try to reestablish the desired
$R_{ext}=1.23$ m, the beam moves outward and the discharges terminate
due to the collision of the beam with the outer limiter. \\
It has to be noted that in the second scenario
the new controller is activated not by the detection of the CQ or current plateau onset, 
as for the first scenario, but because the HXR signal is above 0.2 for more than 10 ms.
This safety condition triggers the new current ramp-down and $R_{ext}$ references
in the shots  $\#38519$ and $\#38513$.   
The decreasing of $R_{ext}$ reference before the plateau onset, although not
clearly visible, might be associated with a larger initial loss of the runway electrons 
in the high-field side of the vessel and this will be take into consideration for future  
controller design. Furthermore, in order to further reduce RE beam interaction with
the inner vessel, we could even consider the DIII-D approach 
activating a saturated control to decrease as fast as possible the vertical 
magnetic field produced by the active coils (F and V) whenever the CQ is detected.
The selection of the saturated control time duration would require detailed analysis on FTU. 
The PCS-REf1/2 results seem to suggest the 
importance of reducing  the external radius reference to 
minimize the RE interaction with the vessel, confirming similar results 
discussed in \cite{REshift}.  As additional evidence observe the rise of the FC 
signal in correspondence of the increase of $R_{ext}$ in the discharges
run with the old controller of Figure \ref{fig:control_spontaneous} (left column).

\begin{figure}[h]
\begin{center}
\resizebox{1.15\columnwidth}{!}{\includegraphics{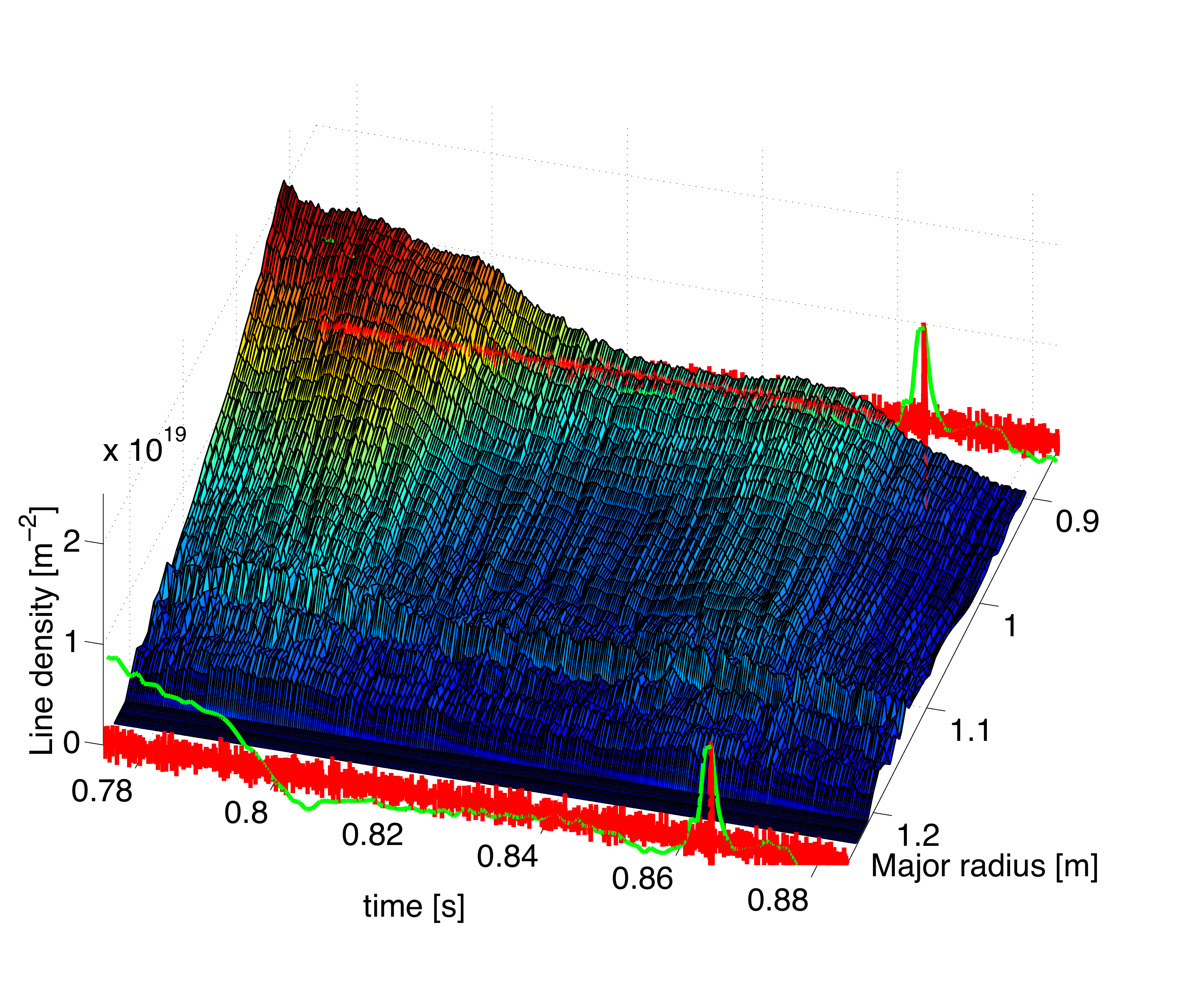}
\includegraphics{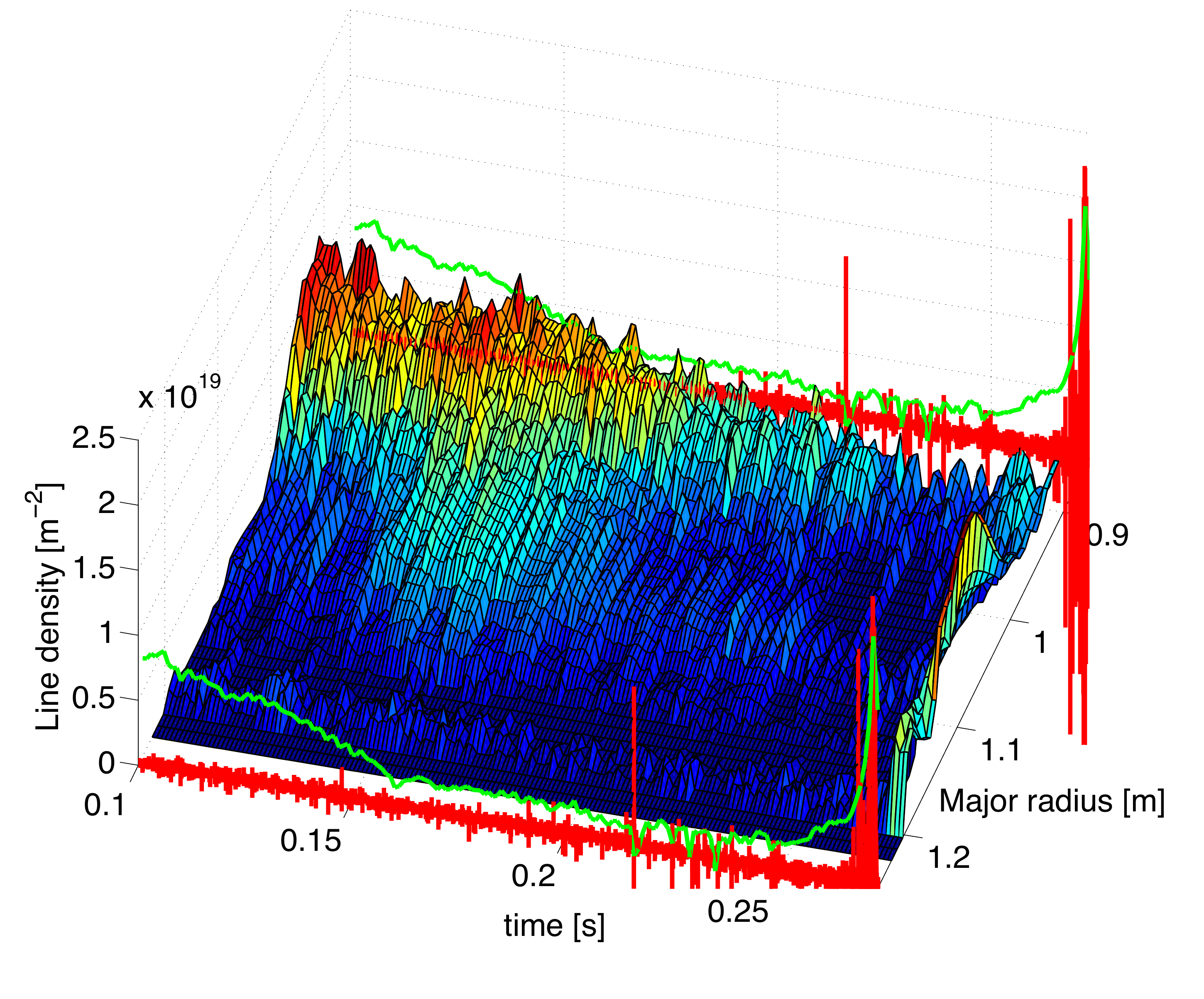}}
\caption{Time evolution the interferometer LOS for RE beam controlled 
discharges $\#36574$ (left) and $\#38519$ (right). The solid red and 
green time traces at major radii 1.25m (duplicated at 0.85m) are the 
MHD  and $V_{\mbox{\tiny loop}}$ signals, respectively, scaled by 1E18.} 
\label{fig:36574_3D} 
\end{center}
\end{figure}

A further interesting feature is that despite the RE beam final current loss in the 
PCS-REf2 (right column of Figure \ref{fig:control_spontaneous}) is larger, 
the corresponding final FC peaks are noticeably smaller if compared with the final peaks 
obtained in Figures \ref{fig:control_neon} and \ref{fig:control_spontaneous}.
It is interesting to note also that in the shot $\#38519$ (slightly in the $\#38513$) 
the HXR signal drops below the saturation threshold before the final loss unlike 
all the other discharges at FTU: since the current drop at final loss is about 100 kA,
this might suggest that a considerable RE beam energy has been dissipated 
during the ramp-down.  
\begin{figure}[h]
\begin{center}
\resizebox{1.15\columnwidth}{!}{\includegraphics{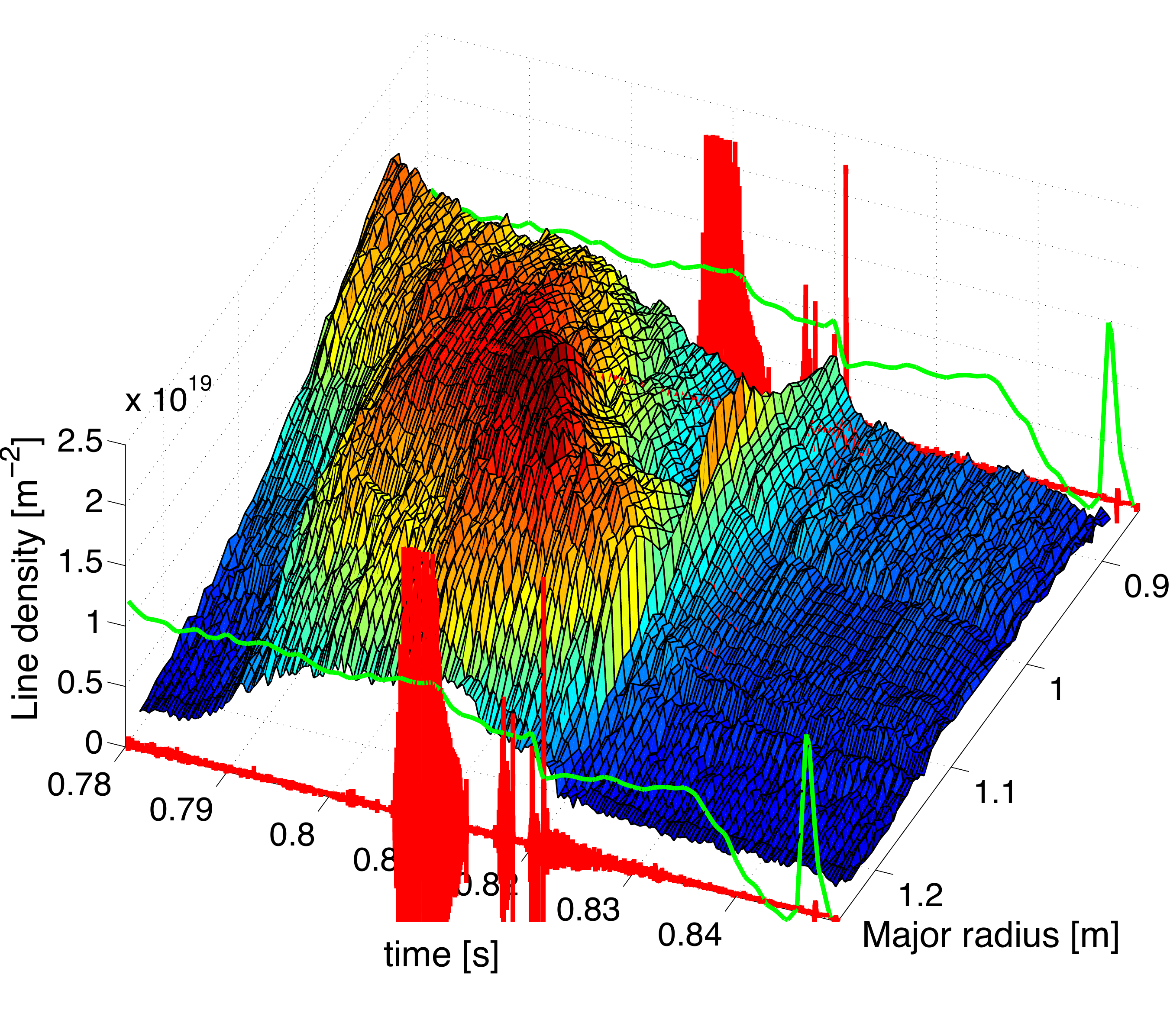}
\includegraphics{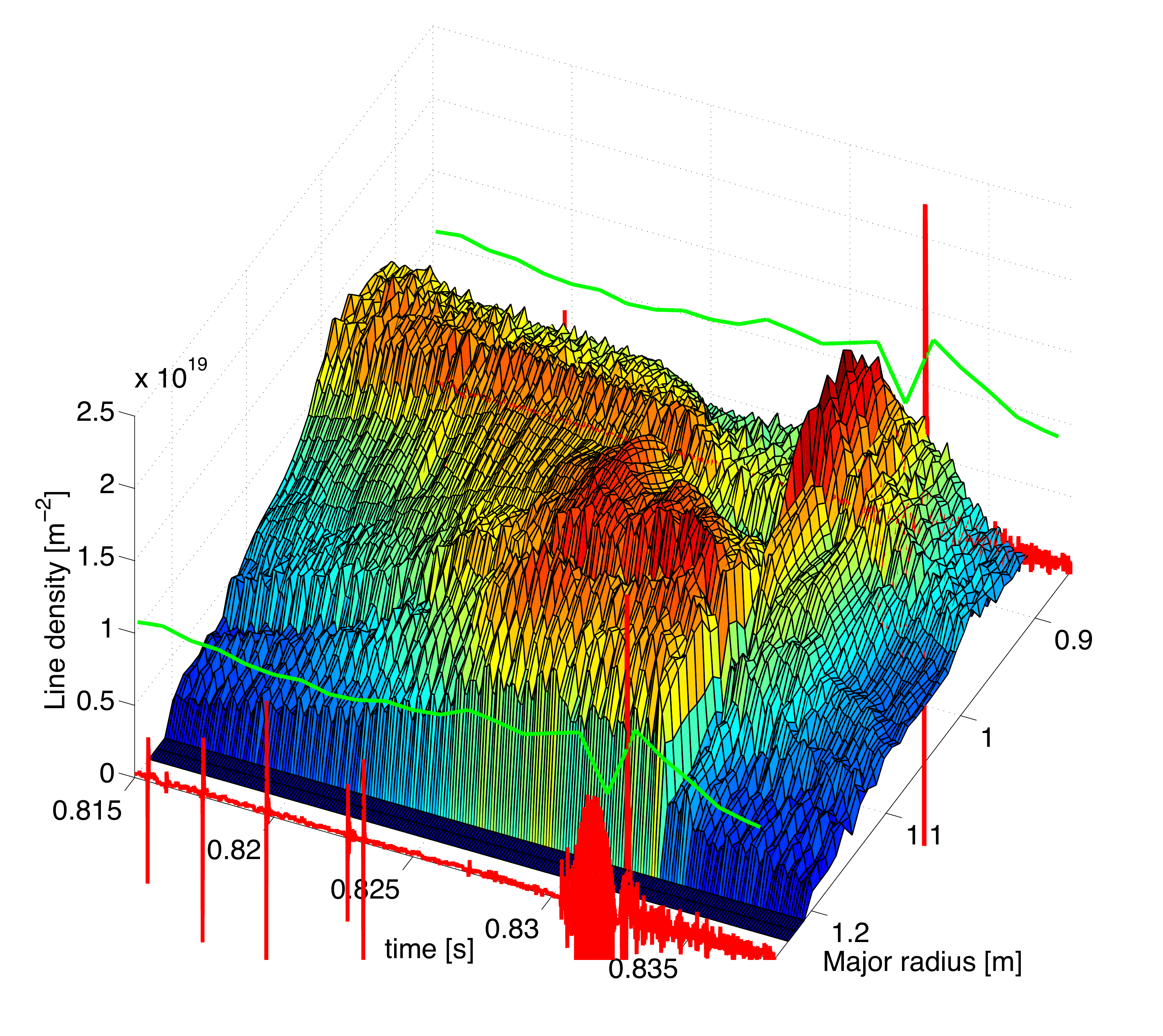}}
\caption{RE beam plateaus without RE control active: Time evolution (left 
$\#35965$ and right $\#35964$) of the interferometer LOS at different major radius. 
The solid red and green time traces at major radii 1.25 m (duplicated at 0.85 m) are the 
MHD and $V_{\mbox{\tiny loop}}$ signals, respectively, scaled by 1E18.} 
\label{fig:35965_3D} 
\end{center}
\end{figure} 
These facts are signs that certain degree of RE energy dissipation 
has been obtained reducing the slope of the $I_p$ ramp-down and 
the $R_{\mathrm{ext}}$ reference.\\
%\dancom{}{Furthermore, it is interesting 
%to note that the HXR signal falls below the saturation threshold even 
%before the final loss, meaning that only few (or with decreased energy) RE survive to 
%the current ramp-down. }
We proceed now to the analysis of the 
time evolution of the interferometer radial profiles.
Figure \ref{fig:35965_3D} shows the time interval between the RE plateau onset 
and the final loss ($I_p\geq 20 kA$),  for discharges $\#35965$ and $\#35964$
obtained with the old controller in the first experimental scenario. 
Unfortunately we do not have the scanning interferometer data 
for the second scenario experiments with the old controller. 
Figure \ref{fig:36574_3D} shows two experiments with the new controllers: 
a) PCS-REf1 (discharge $\#36574$, first plasma scenario) 
b) PCS-REf2 (discharge $\#38519$, second plasma scenario.
Fig. \ref{fig:35965_3D} indicates that the RE plateau 
termination is triggered by MHD instabilities that suddenly move the 
background plasma/RE beam inward. 
Further studies are necessary in order to better understand the 
instability type, able to induce such RE beam displacements, shown in 
Fig. \ref{fig:35965_3D} for discharges $\#35964$ and $\#35965$.
By looking at the Fig. \ref{fig:36574_3D} it is evident that the new control system 
is able to avoid the large outer oscillation shown in Fig. \ref{fig:35965_3D}
as well as in the left columns of Fig. \ref{fig:control_spontaneous} 
and \ref{fig:control_neon} .\\
%It is important to note that the final current loss occurring in the 
%shots ($\#36574$, $\#38519$) of the second scenario 
%are consistently larger than the ones in the first scenario.
%Causes of such different losses are still under investigation: signal
%analysis seems to suggest that premature RE beam loss should be 
%caused by vertical instability, probably related to a large RE beam elongation
%induced by \dancom{high amplitude reached by $I_F$}{the vertical field produced 
%by the coils F and V during ramp-down}. In fact, for standard plasmas \dancom{}{in FTU} the 
%elongation $\hat e$ satisfies $\hat e \approx k_0 - k_1 I_F/I_p$, 
%with $k_0=1.03$ and $k_0=-4.61$. This would suggest to use 
%the Current Allocator \dancom{also}{quickly enough (modifying its convergence
%parameters at run-time when the plateau is detected)} during the RE beam current ramp-down 
%\dancom{}{ to promptly  reduce the $I_F$ current amplitude hence reducing the beam elongation} and 
%it is the subject of future experiments. \\
\begin{figure}[h]
\begin{center}
\resizebox{1.0\columnwidth}{!}{\includegraphics{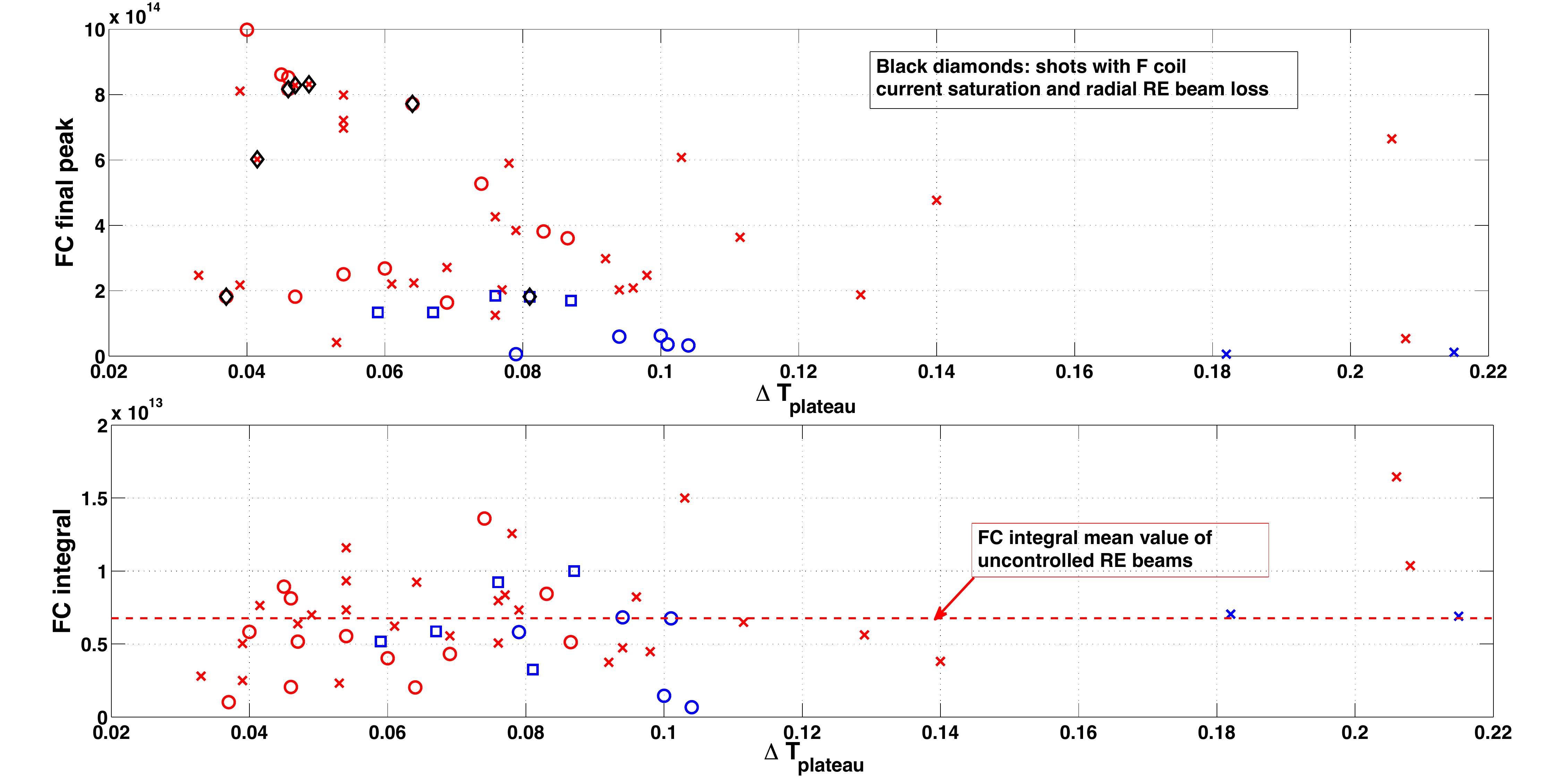}}
\caption{A comparison of different disruption generated 
RE beams. In red circles  and crosses the first and second scenarios RE beam plateaus
with old control policy, respectively. In blue circles and blue crosses the RE beams controlled
by the  PCS-REf1(first scenario) and PCS-REf2 (second scenarios), whereas in 
blue squares are the shots with only a current ramp-down without a redefinition of $R_{ext}$
reference nor the use of the current allocator. An example of the latter type of shot is shown in Fig. 
\ref{fig:control_spontaneous}  ($\#36912$).  A black diamond is superimposed to the shots
where the $I_F$ reached the saturation threshold. } 
\label{fig:AllShots_v2a} 
\end{center}
\end{figure} 

\begin{figure}[h]
\begin{center}
\resizebox{1.0\columnwidth}{!}{\includegraphics{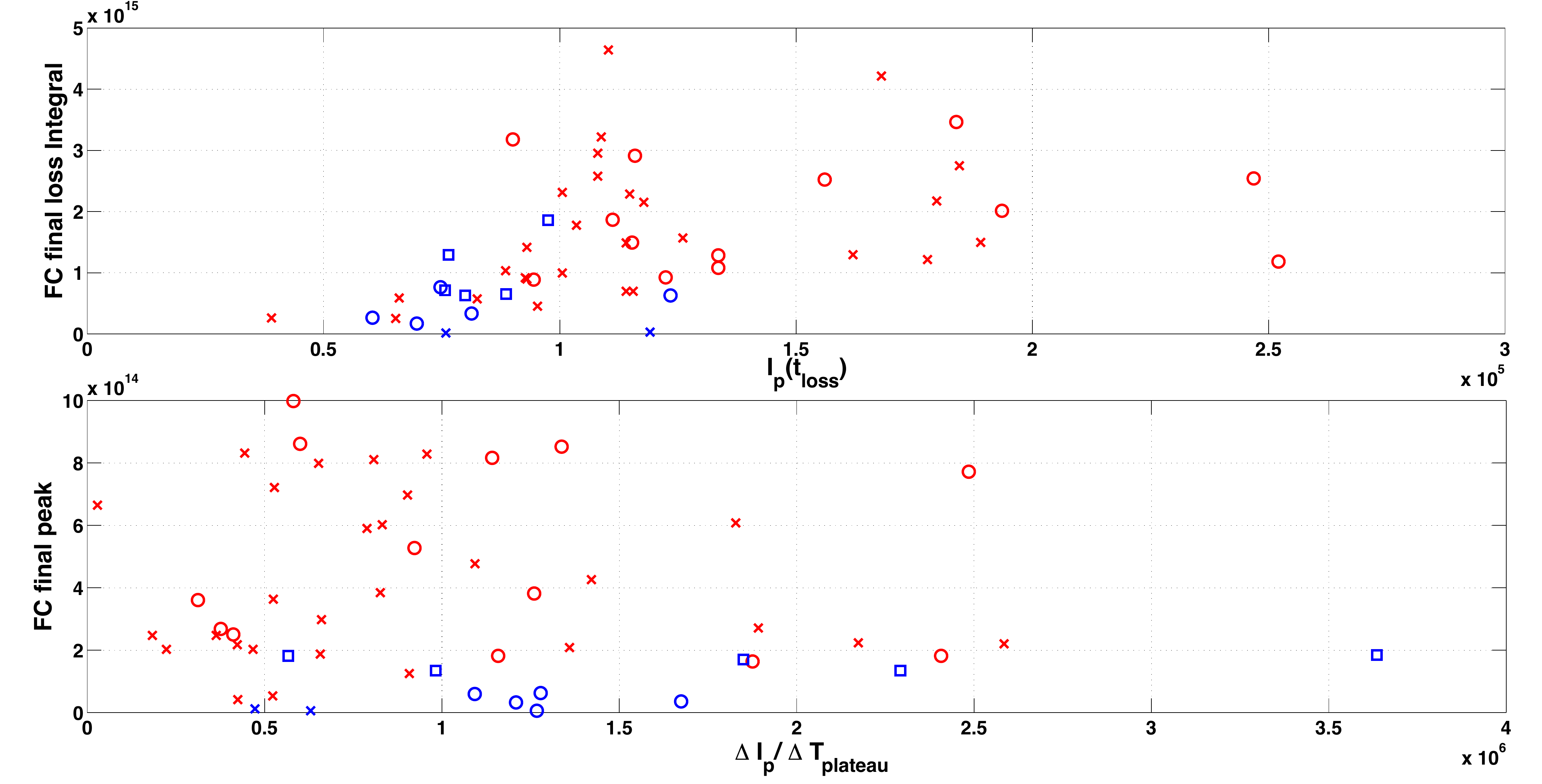}}
\caption{Comparative studies of the FC tail integral  vs $I_p$ at the onset
of the final current loss (top) and the FC final peak vs $I_p$ decaying rate (bottom). } 
\label{fig:AllShots_v2b} 
\end{center}
\end{figure}

\noindent Finally, in Figures \ref{fig:AllShots_v2a} and \ref{fig:AllShots_v2b} 
we show a comparison of 52 disruption generated 
RE beam plateaus, retrieved by ad-hoc algorithm on the FTU database among
35000 discharges, subdivided as follow: 2 shots with PCS-REf2 in the second scenario 
(blue cross), 5 shots with PCS-REf1 in the first scenario (blue circle), 
5 shots with only a linear current ramp-down without $R_{ext}$ redefinition 
nor the current allocator active in the second scenario (blue square) and 
with the old controller 16 shots in the first scenario (red circle) and 24 shots in 
the second scenario (red cross).  A black diamond is superimposed to the shots
where the $I_F$ reached the saturation threshold.\\
The definition  
$\Delta T_{\mbox{\tiny plateau}}=t_{\mbox{ \tiny FCspike}}-t_{CQ}$ is considered, 
where the $t_{CQ}$ is the onset of the current drop and $t_{loss}$
it is assumed to be the time corresponding to the last FC spike before 
$|I_p|$ drops below 40 kA, and generally individuates the knee of the 
$I_p$ time traces at final loss. 
The FC integral is evaluated as the sum of all counts from the beginning of the CQ
up to the end of the shot.\\
In the top plot of the Figure \ref{fig:AllShots_v2a} the FC final peak values are 
shown with respect to the RE beam plateau duration. The values obtained by 
PCS-REf2 are the smallest (blue crosses). Low levels are also obtained by 
PCS-REf2 (blue circles), whereas the only current ramp-down without 
the $R_{ext}$ redefinition and current allocator (blue square) are slightly 
above the former. For the shots in which the current of the coil $I_F$
reaches the saturation, leading to plasma radial loss of confinement, 
a black diamond is superimposed. The saturation of the coil F is not 
observed in the shots where the current allocator, that modifies the 
current $I_V$ in order to maintain $I_F$ far from saturation thresholds, 
is used (PCS-REf1,PCS-REf2). The reduction of the external
radius goes along the same direction of reducing $I_F$ excursions. It
is interesting to note that the final FC peaks of shots with only
a current ramp-down (blue square) are higher than PCS-REf1 and 
PCS-REf2: a possible motivation of this difference, as well as the 
plateau duration, is the redefinition of the $R_{ext}$ reference and the use
of the Curren Allocator. \\
In the bottom plot of the Figure \ref{fig:AllShots_v2a}
the FC integral is shown with respect to the RE beam plateau duration.
In this case, the mean value of blue circles is slightly below the mean 
value of the FC integral evaluated with respect to the uncontrolled 
shots (red cross and circles) meanwhile  the blue crosses have approximatively  
the same value. It is the worth of mentioning that the integral of the FC 
x-ray monitor is proportional to the total energy absorbed by the vessel
during the shot, whereas the FC values are proportional to the power 
released by the RE beam onto the vessel.
Since we are not able to reconstruct the RE beam impact surfaces, we 
can not evaluate the actual power deposition. \\

\begin{figure}[h]
\begin{center}
\resizebox{1.0\columnwidth}{!}{\includegraphics{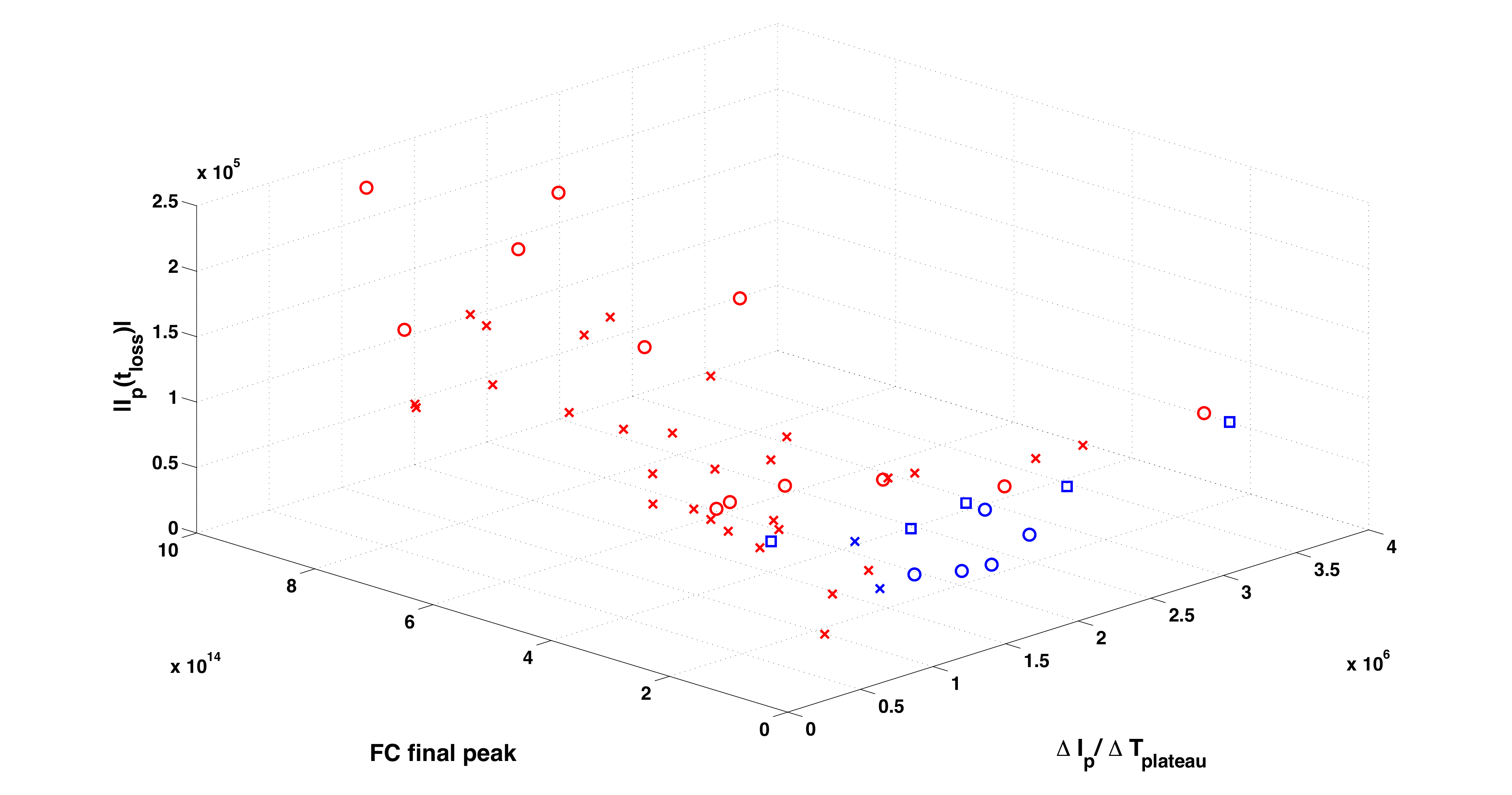}}
\caption{Relation between $|I_p(t_{\mbox{\tiny loss}})|$ on the z-axis,  
FC final peaks on the y-axis and decaying rate on the x-axis for 52 disruption generated 
RE beam.} 
\label{fig:3Dcomare} 
\end{center}
\end{figure}

\noindent
Given that the FC integral mean values of blue circles and crosses
are about the same of red ones, whereas the final FC peaks are much smaller for the
newly controlled RE beam, we could infer that slow current ramp-down and $R_{ext}$
redefinition allow to decrease the RE beam energy and possibly reduce its interaction with the PFC.\\
In the top plot of the Figure \ref{fig:AllShots_v2b} the FC final loss integral,
evaluated summing up all the counts of the FC camera from $t_{\mbox{\tiny loss}}$ to 
the end of the shot versus the $|I_p(t_{\mbox{\tiny loss}})|$ is shown. The onset
of the $I_p$ final loss defined  is estimated as $t_{\mbox{\tiny loss}} = t_{\mbox{ \tiny FCspike}}-5$. 
This figure has been provided since the ratio between the FC tail integral and 
$|I_p(t_{\mbox{\tiny loss}})|$ should be related to percentage of the current
 (energy and number) still carried by the runaways before the final loss onset.\\
In the bottom plot of the same figure,  the FC final peak value is shown
with respect to the decaying rate of $I_p$,  evaluated as the ration between 
$\Delta I_p = |I_p(t_{\mbox{\tiny plateau}})-I_p(t_{\mbox{\tiny loss}})|$ and 
$\Delta T_{\mbox{\tiny plateau}}$, where $t_{\mbox{\tiny plateau}}$ is the 
onset of the RE beam plateau. It has to be noted that from this 2D picture
it is not possible to see the value of $|I_p(t_{\mbox{\tiny loss}})|$ and 
the shots with high FC final peaks that seem to have a small current decaying rate are 
indeed shots with premature loss of confinement, leading to small $\Delta I_p$ 
and than a small decaying rate. To better show this dependence we 
add in the Figure \ref{fig:3Dcomare} the dependence by   
$|I_p(t_{\mbox{\tiny loss}})|$. From these last two pictures it seems clear that to 
have small $|I_p(t_{\mbox{\tiny loss}})|$ and FC peaks the current decaying rate
is below 2MA/s. 

\section{Conclusions}

Two algorithms for the control of
disruption generated RE have been implemented at FTU. The two algorithms 
redefine in real-time the $R_{\mathrm{ext}}$ and $I_p$ ramp-down references, exploiting 
the magnetic and gamma-rays signals. The $I_p$ ramp-down is performed
via the central solenoid and the current in the poloidal coils is changed to
control the position of the RE beam as determined by the magnetic measurements.\\
We have found that the 
external plasma radius $R_{\mathrm{ext}}$ evaluated by magnetic moments 
at FTU can be used to estimate the RE beam radial position when 
the current profiles are not heavily peaked as shown by the interferometer data.
It has been shown that modifying the plasma current reference (ramp-down) 
and reducing the $R_{\mathrm{ext}}$ the gamma signal, provided by the 
FC chamber, decreases as an indication of RE beam energy suppression and reduced interactions 
with the vessel (especially the low-field side). A (slow) current decay rate of about 
0.5 MA/s has been found to provide a better RE beam confinement and consequently 
a controlled energy dissipation. To further and quantitatively corroborate this fact, 
we have analyzed a considerable amount of post-disruption RE beam
discharges at FTU showing that FC peaks at final loss decrease when slowly 
ramped-down. This is in accord with experimental findings in
\cite{REshift}, where slow (1MA/s) current ramp-down have seen to provide better RE beam 
confinement.
Although further experiments are necessary to better 
refine the optimal external radius reference during the RE beam current ramp-down,
possibly defined as a function of $I_p$ and RE beam energy in future work, 
a constant value of approximatively $1.12$ m 
seems to significantly help in reducing the RE impacts with the PFC.
This corresponds to an external major radius reduction of 
approximatively $10 \%$ of the flat-top value ($1.23$ m) and a $40 \%$ reduction 
of to the plasma minor radius that  is equal to 0.305 m in FTU. 
The interferometer signal analysis
has shown that strong MHD induced instabilities, which displace a large percentage
of the RE beam, arise when the (cold) electron profiles are highly peaked
as in \cite{smith_09}.
Now that we have found suitable plasma/RE beam current and position 
references, the current and position controllers PID-T and PID-F 
will be re-designed to further improve their performances specifically in the 
RE control phase. The novel controllers will be based on  a RE beam dynamical model 
identification that is under development.

\section{Acknowledgements}
This work was supported by the EU Horizon 2020 research and innovation
program (Project WP14-MST2-9: Runaway Electron Studies in FTU) - ENEA-EUROFUSION.\\

\end{document}